\documentclass[%
reprint,
 amsmath,amssymb,
 aps,
]{revtex4-2}
\usepackage[letterpaper,top=2cm,bottom=2cm,left=3cm,right=3cm,marginparwidth=1.75cm]{geometry}
\usepackage{graphicx}
\usepackage{dcolumn}
\usepackage[bottom]{footmisc}
\usepackage{bm}

\begin{document}


\title{Van der Waals Gravity Theory}
\author{H. R. Fazlollahi}
\email{seyed.hr.fazlollahi@gmail.com (Corresponding Author)}
\affiliation{%
 PPGCOSMO \& Departamento de Física, Universidade Federal do Espírito Santo (UFES), Av. Fernando Ferrari, 514 Campus de Goiabeiras, Vitória, Espírito Santo CEP 29075-910, Brazil}%

\begin{abstract}

In this study, we propose an extension of general relativity inspired by the van der Waals equation of state, incorporating non-ideal thermodynamic effects into the gravitational sector. Our approach is based on the thermodynamic interpretation of gravity introduced by Jacobson, in which the field equations arise from the Clausius relation. Within this framework, we obtain modified gravitational field equations in which the effective gravitational coupling is no longer constant, but instead evolves with the properties of the underlying spacetime system. This dynamical behavior leads to significant consequences in high-energy regimes. In particular, it provides a natural mechanism for avoiding the initial singularity of standard Big Bang cosmology and gives rise to non-singular black hole solutions. These findings indicate that incorporating non-ideal thermodynamic features into the description of spacetime may offer a consistent route toward resolving fundamental singularities in classical gravitational theory.

\begin{description}
\item[Keywords]
Van der Waals Fluid; Gravity Theory; Non-Singular System.
\end{description}
\end{abstract}

\maketitle

\section{\label{sec:level1}INTRODUCTION}

The formulation of gravity as a manifestation of spacetime curvature by Einstein \cite{Einstein:1915ca} opened a profound and enduring perspective on one of the fundamental interactions in nature. General relativity has achieved remarkable success, from accurately accounting for the perihelion precession of Mercury \cite{Einstein:1915bz} to providing an excellent description of gravitational phenomena within the Solar System \cite{Dyson:1920cwa}. Nevertheless, when gravitational effects are examined on galactic and cosmological scales, particularly in the context of galaxy stability and the large-scale evolution of the Universe, Einstein’s theory encounters significant observational challenges \cite{Toomre:1964zx, Ishak:2018his, Weinberg:2013agg}.

Two prominent inconsistencies arise in this broader regime. At galactic scales, the observed dynamics suggest the presence of an additional, non-luminous mass component, commonly attributed to dark matter \cite{Freese:2008cz, Garrett:2010hd, Hui:2021tkt, Navarro:1995iw, Bertone:2004pz, Jungman:1995df, Arkani-Hamed:2008hhe}. On cosmological scales, independent observations indicate that the Universe is currently undergoing an accelerated expansion phase, often associated with dark energy \cite{Peebles:2002gy, Linder:2002et, Frusciante:2019xia, Planck:2018vyg, WMAP:2003elm, Copeland:2006wr, SupernovaSearchTeam:1998fmf, SupernovaCosmologyProject:1998vns}. These phenomena are not naturally explained within the standard framework of general relativity without introducing new, and so far unobserved, components.

Beyond these observational issues, there exists a fundamental theoretical tension between general relativity and quantum mechanics \cite{Ashtekar:1997yu, Gibbons:1976ue, DeWitt:1967uc}. The former is inherently non-linear, while the latter is formulated as a linear theory. This mathematical incompatibility poses a serious obstacle to constructing a unified framework. A straightforward combination of these theories leads to conceptual and technical difficulties, indicating that such an approach is unlikely to yield a complete and consistent description of fundamental interactions \cite{tHooft:1974toh, Thiemann:2001gmi, Thiemann:2007pyv}.

In response to these challenges, two broad research directions have emerged. One approach seeks a more fundamental theory that transcends both general relativity and quantum mechanics, aiming to unify them within a deeper conceptual framework. Prominent candidates in this direction include string theory \cite{Aharony:1999ti, Gross:1987ar, Witten:1995ex, Polchinski:1998rq} and its various extensions \cite{Becker:2006dvp, Schwarz:1982jn, Zwiebach:2004tj}, which attempt to incorporate both quantum and gravitational phenomena at a fundamental level. Despite their theoretical richness, such models face substantial conceptual and observational challenges \cite{deCarlos:2005kh, Sikivie:2006ni, Banks:2001yp, Susskind:2003kw}.

The alternative approach assumes the validity of quantum mechanics at microscopic scales and instead focuses on modifying the gravitational sector. Within this framework, various extensions of general relativity have been proposed, motivated by both phenomenological considerations \cite{DeFelice:2010aj, Capozziello:2011et, Harko:2011kv} and deeper theoretical insights \cite{Lovelock:1971yv, Brans:1961sx, Rastall:1972swe, Fazlollahi:2023rhg}. These modified gravity theories often succeed in addressing certain observational discrepancies \cite{Clifton:2011jh, Nunes:2016drj} and may provide hints toward a more fundamental connection between gravity and quantum phenomena \cite{Stelle:1976gc, Horava:2009uw, Deser:2007jk}.

An especially intriguing line of investigation suggests that thermodynamics may provide fundamental and reliable clues toward understanding the connection between quantum mechanics and general relativity. Thermodynamics constitutes a powerful and universal framework, capable of describing a wide range of physical systems, and has proven to be deeply consistent with several aspects of quantum theory~\cite{Skrzypczyk:2016lnb, Bera:2018ozi, Deffner:2019iai}. On the other hand, thermodynamic principles can also be employed to derive the dynamical laws of gravity. A pioneering contribution in this direction was made by Jacobson, who demonstrated that Einstein’s field equations can be obtained from the Clausius relation,
\begin{equation}\label{eq:1.1}
    \delta Q = T\, dS,
\end{equation}
by applying it to local Rindler horizons \cite{Jacobson:1995ab}. In this approach, gravity does not emerge as a fundamental interaction, but rather as an effective, thermodynamic description of spacetime. In this formulation, $\delta Q$ represents the energy flux across the horizon, while $T$ corresponds to the Unruh temperature perceived by an accelerated observer. This result suggests that the gravitational field equations may be interpreted as thermodynamic equations of state of spacetime itself.

Following this insight, extensive work has been carried out to explore the deep connections between gravitational dynamics and thermodynamic principles \cite{Padmanabhan:2003gd, Eling:2006aw, Padmanabhan:2009vy}. In particular, it has been shown that, in a Friedmann-Robertson-Walker (FRW) Universe, the Friedmann equations can be recast in the form of the first law of thermodynamics on the apparent horizon \cite{Cai:2005ra, Akbar:2006er}. Such formulations have been applied to various entropy models and have provided useful phenomenological descriptions of the late-time acceleration of the Universe \cite{Sheykhi:2018dpn, Sheykhi:2022jqq, Pasqua:2015bfz, Fazlollahi:2022bgf}. However, many of these approaches operate at the level of cosmological dynamics, typically through the Friedmann equations, rather than arising from a deeper reformulation of general relativity itself. This naturally raises the question of whether one can instead revise the underlying gravitational framework from which such equations emerge.

In this regard, a closer examination of Jacobson’s formalism reveals that the gravitational field equations follow from the Clausius relation \eqref{eq:1.1} under assumptions analogous to an ideal thermodynamic system. Yet, real physical systems are more accurately described by the van der Waals equation of state, which accounts for intermolecular interactions and finite-size effects. This insight motivates the investigation of whether adopting a van der Waals–type structure at the fundamental level can lead directly to a more general form of the gravitational field equations. We therefore pursue this idea by extending the thermodynamic derivation of gravity beyond the idealized framework. Specifically, we replace the implicit ideal gas behavior with a van der Waals-type equation of state and derive the corresponding modifications to the gravitational field equations within the Jacobson formalism. 

The remainder of this study is organized as follows. Section II briefly reviews the thermodynamic interpretation of gravity and establishes the necessary formalism. In Section III, the Clausius relation is modified by incorporating the van der Waals equation of state, and the corresponding field equations are derived. In Section IV, the implications of this framework are examined in two distinct settings: first, within a cosmological context by applying it to the Friedmann-Robertson-Walker Universe, and second, through the analysis of spherically symmetric solutions, including possible black hole configurations. Finally, the main results are summarized in Section V.

Throughout this work, we adopt natural units $\hbar = c = 1$. Following standard notation, $dX$ denotes an infinitesimal change in a variable $X$, while $\delta X$ represents a variation of $X$ \cite{Jacobson:1995ab}.

\section{Revisiting Jacobson’s Proposal}

The Einstein field equations may be understood as a consequence of a local thermodynamic condition imposed on spacetime, rather than as fundamental dynamical laws. In this context, Jacobson formulated the idea that, at every spacetime point $p$, the Clausius relation \ref{eq:1.1} holds for all local Rindler horizons passing through $p$ \cite{Jacobson:1995ab}.

To give a precise meaning to this statement, consider an arbitrary point $p$ in spacetime and construct, in its neighborhood, a local causal horizon generated by a congruence of null geodesics. The generators of this congruence are described by a future-directed null vector field $k^\mu$, affinely parametrized by $\lambda$, such that $k^\mu = \mathrm{d}x^\mu / \mathrm{d}\lambda$. Associated with this local horizon is an approximate boost Killing vector field $\chi^\mu$, which vanishes at the point $p$ and generates translations along the horizon. In the immediate vicinity of $p$, this vector field can be expressed as
\begin{equation}\label{eq:2.1}
    \chi^\mu = -\kappa \lambda k^\mu,
\end{equation}
where $\kappa$ is the surface gravity characterizing the acceleration of the local Rindler observer.

Within this framework, the energy flux $\delta Q$ through the horizon is defined as \cite{Jacobson:1995ab}
\begin{equation}\label{eq:2.2}
    \delta Q = \int T_{\mu\nu} \, \chi^\mu \, \mathrm{d}\Sigma^\nu,
\end{equation}
where $T_{\mu\nu}$ is the energy--momentum tensor, and $\mathrm{d}\Sigma^\nu = k^\nu \, \mathrm{d}\lambda \, \mathrm{d}A$ is the natural surface element on the null hypersurface, with $\mathrm{d}A$ denoting the area element of a spatial cross-section of the horizon. Substituting the expression for $\chi^\mu$, one obtains
\begin{equation}\label{eq:2.3}
    \delta Q = - \int \kappa \lambda \, T_{\mu\nu} \, k^\mu k^\nu \, \mathrm{d}\lambda \, \mathrm{d}A.
\end{equation}
Since the construction is local and restricted to an infinitesimal neighbourhood around $p$, the surface gravity $\kappa$ may be treated as constant, yielding
\begin{equation}\label{eq:2.4}
    \delta Q = - \kappa \int \lambda \, T_{\mu\nu} \, k^\mu k^\nu \, \mathrm{d}\lambda \, \mathrm{d}A.
\end{equation}
This relation shows that the heat flux is governed by the projection $T_{\mu\nu} k^\mu k^\nu$, which measures the flow of energy along the null generators.

The temperature is identified with the Unruh temperature,
\begin{equation}\label{eq:2.5}
    T = \frac{\kappa}{2\pi},
\end{equation}
while the entropy is assumed to obey the area law,
\begin{equation}\label{eq:2.6}
    \mathrm{d}S = \eta \, \delta A,
\end{equation}
where $\eta = 1/(4G)$ is a universal constant and $\delta A$ denotes the variation of the horizon area. The change in area is determined by the expansion of the horizon generators $\theta$, such that \cite{Jacobson:1995ab}
\begin{equation}\label{eq:2.7}
    \delta A = \int \theta \, \mathrm{d}\lambda \, \mathrm{d}A,
\end{equation}
which measures the fractional rate of change of the cross-sectional area along the null generators.

The evolution of $\theta$ is governed by the Raychaudhuri equation \cite{Raychaudhuri:1953yv},
\begin{equation}\label{eq:2.8}
    \frac{\mathrm{d}\theta}{\mathrm{d}\lambda} = -\frac{1}{2}\theta^2 - \sigma_{\mu\nu}\sigma^{\mu\nu} - R_{\mu\nu} k^\mu k^\nu,
\end{equation}
where $\sigma_{\mu\nu}$ is the shear tensor and $R_{\mu\nu}$ is the Ricci tensor. By choosing the congruence such that, at the point $p$, locally, $\theta = 0$ and $\sigma_{\mu\nu} = 0$, one finds to leading order in $\lambda$,
\begin{equation}\label{eq:2.9}
    \theta \approx - \lambda R_{\mu\nu} k^\mu k^\nu.
\end{equation}
It follows that the change in the area element is
\begin{equation}\label{eq:2.10}
    \delta A = - \int \lambda R_{\mu\nu} k^\mu k^\nu \, \mathrm{d}\lambda \, \mathrm{d}A,
\end{equation}
so that the entropy variation becomes
\begin{equation}\label{eq:2.11}
    \mathrm{d}S = - \eta \int \lambda R_{\mu\nu} k^\mu k^\nu \, \mathrm{d}\lambda \, \mathrm{d}A.
\end{equation}
Substituting the expressions for $\delta Q$, $T$, and $\mathrm{d}S$ into the Clausius relation, one obtains
\begin{equation}\label{eq:2.12}
    \int \lambda R_{\mu\nu} k^\mu k^\nu \, \mathrm{d}\lambda \, \mathrm{d}A = 8\pi G \int T_{\mu\nu} \lambda k^\mu k^\nu \, \mathrm{d}\lambda \, \mathrm{d}A.
\end{equation}
Since this equality must hold for every infinitesimal horizon patch and for all null vectors $k^\mu$ at the point $p$, It follows that the difference of the integrands is necessarily proportional to the metric tensor. Hence,
\begin{equation}\label{eq:2.13}
    R_{\mu\nu} - 8\pi G T_{\mu\nu} = f \, g_{\mu\nu},
\end{equation}
where $f$ is a scalar function. Imposing local conservation of energy-momentum, $\nabla^\mu T_{\mu\nu} = 0$, together with the contracted Bianchi identity, one finds
\begin{equation}\label{eq:2.14}
    f = \frac{1}{2} R - \Lambda,
\end{equation}
where $\Lambda$ is a constant. This leads directly to the Einstein field equations \cite{Jacobson:1995ab},
\begin{equation}\label{eq:2.15}
    R_{\mu\nu} - \frac{1}{2} R g_{\mu\nu} + \Lambda g_{\mu\nu}
= 8\pi G T_{\mu\nu}.
\end{equation}
This derivation demonstrates that the gravitational field equations can emerge from a purely thermodynamic argument, provided that the Clausius relation is assumed to hold locally. In particular, Jacobson’s construction implicitly relies on an equilibrium thermodynamic description analogous to that of an ideal system. 

In the next section, we go beyond this idealized assumption and replace the underlying thermodynamic structure with a van der Waals–type equation of state. By constructing a generalized Clausius relation within this framework and following the same thermodynamical procedure, we derive a modified form of the gravitational field equations \eqref{eq:2.15} that captures these non-ideal effects at a fundamental level.

\section{Van der Waals Gravity}

The van der Waals equation provides a fundamental modification of the ideal gas law by incorporating the finite size of molecules and intermolecular interactions. While the ideal gas equation $PV = RT$ assumes non-interacting, point-like particles, real gases exhibit deviations from this behavior, particularly at high pressures and low temperatures \cite{Vovchenko:2015vxa, Vovchenko:2017cbu}. The van der Waals equation introduces two corrective terms and can be written as
\begin{equation}\label{eq:3.1}
    \left(P + \frac{a}{V^2}\right)(V - b) = RT,
\end{equation}
where $a$ characterizes the strength of intermolecular attractive forces, and $b$ represents the excluded volume associated with the finite size of the constituents.

To construct a modified Clausius relation based on this equation of state, we begin with the first law of thermodynamics,
\begin{equation}\label{eq:3.2}
    \delta Q = \mathrm{d}U + P\, \mathrm{\delta}V.
\end{equation}
Assuming that the internal energy is a function of volume and temperature, $U = U(V, T)$, one may use the thermodynamic identity
\begin{equation}\label{eq:3.3}
    \left(\frac{\partial U}{\partial V}\right)_T
= T \left(\frac{\partial P}{\partial T}\right)_V - P.
\end{equation}
Furthermore, the total differential of the internal energy reads
\begin{equation}\label{eq:3.4}
    \mathrm{d}U = \left(\frac{\partial U}{\partial V}\right)_T \mathrm{\delta}V
+ \left(\frac{\partial U}{\partial T}\right)_V \mathrm{d}T.
\end{equation}
Using the van der Waals equation of state into \eqref{eq:3.3} and \eqref{eq:3.4}, one finds
\begin{equation}\label{eq:3.5}
    \mathrm{d}U = \frac{a}{V^2} \, \mathrm{\delta}V
+ \left(\frac{\partial U}{\partial T}\right)_V \mathrm{d}T.
\end{equation}
Inserting this result into the first law yields
\begin{equation}\label{eq:3.6}
    \delta Q = \left(\frac{\partial U}{\partial T}\right)_V \mathrm{d}T + \frac{R}{V - b} \, \mathrm{\delta}V.
\end{equation}
Defining the heat capacity at constant volume as $C_V = (\partial U/\partial T)_V$ and using $\delta Q = T \mathrm{d}S$, one obtains the entropy variation
\begin{equation}\label{eq:3.7}
    \mathrm{d}S = \frac{C_V}{T} \, \mathrm{d}T + \frac{R}{V - b} \, \mathrm{\delta}V.
\end{equation}

This relation represents a modified form of the Clausius relation, incorporating corrections inspired by a van der Waals–type equation of state. Nevertheless, such a formulation cannot be directly implemented within the framework of Ted Jacobson’s proposal. To establish a meaningful correspondence with gravity, a careful reinterpretation of the thermodynamic quantities is required. In the context of general relativity, there is no underlying microscopic description in terms of molecular degrees of freedom. Consequently, the quantity $C_V$ should not be understood as a conventional heat capacity. Rather, within a local equilibrium approximation, the term $C_V \mathrm{d}T$ can be reinterpreted as representing an effective energy flux across the local causal horizon.
\begin{equation}\label{eq:3.8}
    C_V \mathrm{d}T \;\longrightarrow\; \delta Q_{\mathrm{geom}}.
\end{equation}
Moreover, gravitational dynamics are naturally expressed in terms of horizon areas rather than volumes. For a spherical configuration,
\begin{equation}\label{eq:3.9}
    V = \frac{4}{3}\pi r^3, \qquad A = 4\pi r^2,
\end{equation}
which implies
\begin{equation}\label{eq:3.10}
    V = \frac{r}{3} A, \qquad \mathrm{\delta}V = \frac{1} {3}\left(A\, \mathrm{\delta}r + r\, \mathrm{\delta}A\right).
\end{equation}
In the local limit $\mathrm{\delta}r \ll r$, one finds
\begin{equation}\label{eq:3.11}
    R\, \mathrm{\delta}V \simeq \frac{\alpha \, r}{3} \, \mathrm{\delta}A,
\end{equation}
where $\alpha$ encodes the effective contribution of the gas constant $R$ in the gravitational context. Furthermore, in following \eqref{eq:3.10}, the parameter $b$ is reinterpreted as a fundamental volume scale, $b = \frac{r}{3} A_0$, where $A_0$ is a constant reference area. Consequently, we arrive at
\begin{equation}\label{eq:3.12}
    \frac{R\, \mathrm{\delta}V}{V - b} \simeq \frac{\alpha\, \mathrm{\delta}A}{A - A_0} = \alpha \, \mathrm{\delta} \ln (A - A_0).
\end{equation}
Combining these results, the modified Clausius takes the form
\begin{equation}\label{eq:3.13}
\mathrm{d}S = \frac{\delta Q_{\mathrm{geom}}}{T}
+ \alpha \, \mathrm{\delta} \ln (A - A_0).
\end{equation}
In the limit $\alpha \to 0$, the additional contribution vanishes and the standard Clausius relation is recovered.

We now implement Jacobson’s construction using this modified relation. The $\delta Q_{\mathrm{geom}}$ and $dS$ are given by
\begin{align}
\delta Q_{\mathrm{geom}} &= -\kappa \int \lambda \, T_{\mu\nu} k^\mu k^\nu \, \mathrm{d}\lambda \, \mathrm{d}A, \\
\mathrm{d}S &= -\eta \int \lambda \, R_{\mu\nu} k^\mu k^\nu \, \mathrm{d}\lambda \, \mathrm{d}A,
\end{align}
while the additional term becomes
\begin{equation}\label{eq:3.16}
    \alpha \, \mathrm{\delta} \ln (A - A_0) = - \frac{\alpha}{A - A_0} \int \lambda \, R_{\mu\nu} k^\mu k^\nu \, \mathrm{d}\lambda \, \mathrm{d}A,
\end{equation}
Since and follwoing Jacobson's proposal, we have considered infenitesimal area around arbtrary point $p$, which imples $A-A_0\approx const.$. With this assumption for small patch, the above relation can be given, approximately as

\begin{equation}\label{eq:3.17}
    \alpha \, \mathrm{\delta} \ln (A - A_0) = - \alpha\int \frac{\lambda R_{\mu\nu}}{A - A_0}  \,  k^\mu k^\nu \, \mathrm{d}\lambda \, \mathrm{d}A,
\end{equation}
Substituting these expressions into the modified Clausius relation yields
\begin{equation}\label{eq:3.18}
\begin{aligned}
-\frac{1}{4G} \int \lambda R_{\mu\nu} k^\mu k^\nu 
&= -\frac{\kappa}{T} \int T_{\mu\nu} \lambda k^\mu k^\nu \\
&\quad - \alpha \int \frac{\lambda R_{\mu\nu}}{A - A_0}  \,  k^\mu k^\nu \, \mathrm{d}\lambda \, \mathrm{d}A \, .
\end{aligned}
\end{equation}
Using $T = \kappa / (2\pi)$ and rearranging, one obtains
\begin{equation}\label{eq:3.19}
    \int \left[ R_{\mu\nu} - \frac{8\pi G (A - A_0)}{A - A_0 - 4\alpha G} \, T_{\mu\nu} \right] \lambda k^\mu k^\nu \, \mathrm{d}\lambda \, \mathrm{d}A = 0.
\end{equation}
Since this relation holds for all null vectors, it follows that
\begin{equation}\label{eq:3.20}
R_{\mu\nu}
- \frac{8\pi G (A - A_0)}{A - A_0 - 4\alpha G}
\, T_{\mu\nu}
= f\, g_{\mu\nu}.
\end{equation}
Here, instead of imposing local conservation on the ordinary energy-momentum tensor $T_{\mu\nu}$, one requires conservation of an effective energy-momentum tensor, namely
\begin{equation}\label{eq:3.21}
    T_{\mu\nu}^{\mathrm{tot}} = \frac{A - A_0}{A - A_0 - 4\alpha G} \, T_{\mu\nu}, \qquad \nabla^\mu T_{\mu\nu}^{\mathrm{tot}} = 0.
\end{equation}
This leads to the modified field equations
\begin{equation}\label{eq:3.22}
G_{\mu\nu} + \Lambda g_{\mu\nu}
= \frac{8\pi G (A - A_0)}{A - A_0 - 4\alpha G}
\, T_{\mu\nu}.
\end{equation}
This result demonstrates that incorporating a non-ideal thermodynamic structure at the fundamental level naturally leads to a modified gravitational coupling,
\begin{equation}\label{eq:3.23}
    G = \text{const.} \;\longrightarrow\; G_{\mathrm{eff}} = \frac{G (A - A_0)}{A - A_0 - 4 \alpha G}.
\end{equation}
Although variations of an effective gravitational constant have been explored extensively in the literature \cite{Uzan:2010pm, Bonanno:2004ki, Christodoulou:2018xxw}, the present formulation displays two key distinctions. First, $G_{\mathrm{eff}}$ arises intrinsically from the thermodynamic structure associated with non-ideal equations of state, rather than being introduced at a phenomenological level. Second, in contrast to many analogous constructions, $G_{\mathrm{eff}}$ depends explicitly on the surface area and spatial geometry, without involving temporal components \cite{Uzan:2010pm, Bonanno:2004ki, Christodoulou:2018xxw}.

As a result, both $G_{\mathrm{eff}}$ and the associated modified field equations are closely linked to the Holographic Principle \cite{tHooft:1999rgb, Bousso:2002ju} and black hole thermodynamics \cite{Bekenstein:1973ur, Hawking:1982dh}, highlighting the profound interplay between spacetime geometry, thermodynamic structure, and gravitational dynamics. Hence, this framework provides a compelling avenue for connecting gravitational behavior with fundamental thermodynamic and holographic principles.

\section{Cosmology and Black Hole Solutions}

In this section, with the discussion organized into two subsections, we explore how the effective gravitational parameter $G_{\mathrm{eff}}$ offers novel perspectives on cosmic evolution and black hole structure.

\subsection{FRW Universe}

In order to explore the implications of the gravitational theory given by \eqref{eq:3.22}, we consider a homogeneous and isotropic Universe described by a spatially flat Friedmann--Robertson--Walker (FRW) metric. Accordingly, the first Friedmann equation takes the form (setting $\Lambda = 0$ to isolate the pure effects of $G_{\mathrm{eff}}$)
\begin{equation}\label{eq:4.1}
    H^2 = \frac{8\pi G (A - A_0)}{3(A - A_0 - 4\alpha G)} \, \rho,
\end{equation}
where $\rho$ denotes the total energy density of the cosmic fluid and $H = \dot{a}/a$ is the Hubble parameter, with $a(t)$ the scale factor.

For an FRW Universe, the apparent horizon radius is given by $r = H^{-1}$ \cite{Melia:2011fj}, and hence the corresponding area becomes $A = 4\pi H^{-2}$. Substituting this relation into \eqref{eq:4.1}, one obtains
\begin{equation}\label{eq:4.2}
    H^2 = \frac{2\pi \left(3 + 2A_0 G \rho \pm \sqrt{\Delta} \right)}{3(A_0 + 4\alpha G)}, 
\end{equation}
for
\begin{equation}\label{eq:4.3}
    \Delta = (2A_0 G \rho - 3)^2 - 96 \alpha G^2 \rho.
\end{equation}
In viable cosmological models, the late-time Universe must reproduce the standard behavior $3H^2 \simeq 8\pi G \rho$~\cite{Copeland:2006wr, Amendola:2015ksp}. Imposing this requirement selects the negative branch of \eqref{eq:4.2}, leading to
\begin{equation}\label{eq:4.4}
    H^2 = \frac{2\pi \left(3 + 2A_0 G \rho - \sqrt{\Delta} \right)}{3(A_0 + 4\alpha G)}.
\end{equation}
In the early-Universe limit, characterized by a high-density regime $\rho \gg 1$, the dominant contribution to $\Delta$ yields $\Delta \approx (2A_0 G \rho)^2$. It then follows that
\begin{equation}\label{eq:4.5}
    H^2_{\mathrm{early}} \approx \frac{2\pi}{A_0 + 4\alpha G}.
\end{equation}
This result indicates that, in the very early Universe, the Hubble parameter approaches a constant value, corresponding to a de Sitter–like phase with exponential expansion,
\begin{equation}\label{eq:4.6}
    a(t) \sim e^{H_{\mathrm{early}} t}.
\end{equation}
Consequently, the model naturally generates an inflationary phase without the need to introduce additional scalar degrees of freedom. Moreover, in the high-density regime, the Hubble parameter remains finite, signaling that the cosmological evolution is free from the conventional initial singularity. This non-singular behavior originates from the parameter $A_0$, which encodes finite-size corrections in the van der Waals equation of state. Within the gravitational context, $A_0$ acquires a clear geometric interpretation as a minimal, non-vanishing area scale, which motivates the identification
\begin{equation}\label{eq:4.7}
    A_{\mathrm{min}} = A_0 = 4\pi r_{\mathrm{min}}^2 
    \;\longrightarrow\; 
    r_{\mathrm{min}} = \sqrt{\frac{A_0}{4\pi}},
\end{equation}
which can be regarded as a geometric ultraviolet cutoff. As $A \rightarrow A_0$, the effective gravitational coupling satisfies $G_{\mathrm{eff}} \rightarrow 0$, indicating that gravity effectively switches off at very small scales. Consequently, even in the limit $\rho \rightarrow \infty$, the Hubble parameter remains finite, thereby resolving the Big Bang singularity at a geometric level.

In the late times, by contrast, as the Universe expands and $\rho \rightarrow 0$, an expansion of Eq. \eqref{eq:4.4} yields
\begin{equation}\label{eq:4.8}
    H^2 \approx \frac{8\pi G \rho}{3} \left(1+\frac{\rho}{\rho_0}\right), \quad\quad \rho_0=\frac{3}{8\alpha G^2}
\end{equation}
This expression coincides with the form obtained in loop quantum cosmology~\cite{Ashtekar:2011ni, Bojowald:2005epg, Agullo:2016tjh}, indicating a close connection between the present framework and quantum gravitational corrections. In this sense, the corrections emerging from the van der Waals thermodynamic structure naturally reproduce the leading-order quantum modifications, suggesting that the latter may arise from a deeper thermodynamic origin.

\subsection{Black Hole Solution}

To investigate black hole solutions within this framework, we consider a static and spherically symmetric spacetime. Restricting attention to the simplest configuration, namely, an uncharged and non-rotating geometry, the line element can be written as
\begin{equation}\label{eq:4.9}
    ds^2 = -f(r)\, dt^2 + \frac{1}{f(r)}\, dr^2 + r^2 \left(d\theta^2 + \sin^2\theta \, d\phi^2 \right).
\end{equation}
The behavior of the metric function $f(r)$ is governed by the scale dependence of the effective gravitational coupling $G_{\mathrm{eff}}$. In the asymptotic regime, where the horizon area satisfies $A \gg A_0$, one has $G_{\mathrm{eff}} \to G$, and the field equations \eqref{eq:3.22} reduce to those of standard general relativity. Consequently, the solution approaches the Schwarzschild form,
\begin{equation}\label{eq:4.10}
    f(r) \longrightarrow 1 - \frac{2GM}{r}.
\end{equation}
This ensures that, far from the black hole, the theory consistently reproduces the classical weak-field limit.

In contrast, approaching the central region corresponds to the limit $A \to A_0$, or equivalently $r \to r_{\mathrm{min}}$ as defined in \eqref{eq:4.7}. In this regime, the effective gravitational coupling vanishes, $G_{\mathrm{eff}} \to 0$, indicating that gravitational interactions are effectively switched off at very small scales. As a consequence, the spacetime tends towards a regular, locally flat geometry near the core,
\begin{equation}\label{eq:4.11}
    f(r) \longrightarrow 1.
\end{equation}
This behavior signals the absence of a curvature singularity at the center. The existence of a minimal radius $r_{\mathrm{min}}$ provides a natural geometric cutoff, leading to a regular core structure analogous to that encountered in the cosmological sector.

A further consistency requirement is that curvature invariants remain finite throughout the spacetime. In particular, the Kretschmann scalar,
\begin{equation}\label{eq:4.12}
    K = R_{\alpha\beta\gamma\delta} R^{\alpha\beta\gamma\delta}
    = f''(r)^2 + \frac{4}{r^2} f'(r)^2 + \frac{4}{r^4} \big(f(r) - 1\big)^2,
\end{equation}
must remain finite in the vicinity of $r \sim r_{\mathrm{min}}$. This condition guarantees the regularity of the geometry and excludes curvature divergences.

Collecting these requirements, the metric function $f(r)$ must satisfy
\begin{equation}\label{eq:4.13}
\begin{aligned}
    &\text{(i)} \quad A \gg A_0 \;\Rightarrow\; G_{\mathrm{eff}} \to G, \quad f(r) \to 1 - \frac{2GM}{r}, \\
    &\text{(ii)} \quad A \to A_0 \;\Rightarrow\; G_{\mathrm{eff}} \to 0, \quad f(r) \to 1, \\
    &\text{(iii)} \quad A < A_0 \;\Rightarrow\; \text{no physical region (geometric cutoff)}, \\
    &\text{(iv)} \quad f(r), \; f'(r), \; f''(r) \;\text{remain finite near } r \sim r_{\mathrm{min}}.
\end{aligned}
\end{equation}
To enforce these conditions, we introduce the phenomenological ansatz
\begin{equation}\label{eq:4.14}
    f(r) = 1 - \frac{2GM \, (r - r_0)^{n/2}}{(r + r_0)^l},
\end{equation}
where the parameters $n$ and $l$, and $r_0$ are chosen to ensure the required asymptotic behavior and regularity. This form provides a smooth interpolation between the Schwarzschild solution at large radii and a regular core at small scales.
Condition (i) immediately yields
\begin{equation}\label{eq:4.15}
    l - \frac{n}{2} = 1,
\end{equation}
while condition (ii) requires $n > 0$. Condition (iii), together with the requirement of a smooth termination of the geometry at $r = r_0$, suggests that $n$ should be an odd integer. We therefore parametrize $n = 2N + 1$, with $N \in \mathbb{N}$. Finally, condition (iv), ensuring the finiteness of curvature invariants, imposes $N \geq 2$.

Choosing the minimal admissible value $N = 2$, one obtains
\begin{equation}\label{eq:4.16}
    n = 5, \qquad l = \frac{7}{2}.
\end{equation}
Upon substituting these values into Eq. \eqref{eq:4.14}, we obtain
\begin{equation}\label{eq:4.17}
    f(r) = 1 - 2GM \sqrt{\frac{(r - r_0)^5}{(r + r_0)^7}}.
\end{equation}
By construction, this metric function satisfies all the conditions listed above and thus represents a consistent regular black hole solution within the present framework.

\begin{figure}[!h]
    \includegraphics[width=0.45\textwidth]{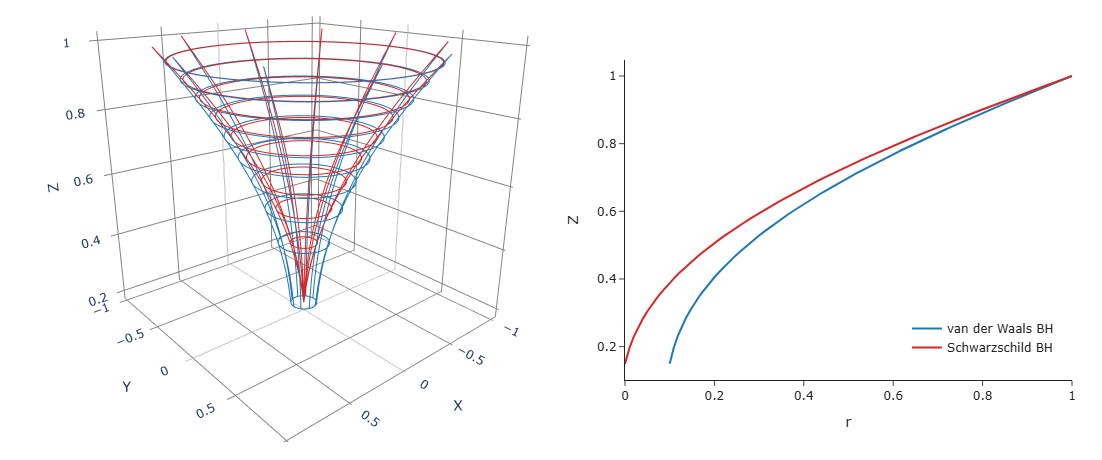}
    \caption{Embedding diagram of the spatial geometry compared with the Schwarzschild case.}
    \label{fig:1}
\end{figure}

\begin{figure}[!h]
    \includegraphics[width=0.45\textwidth]{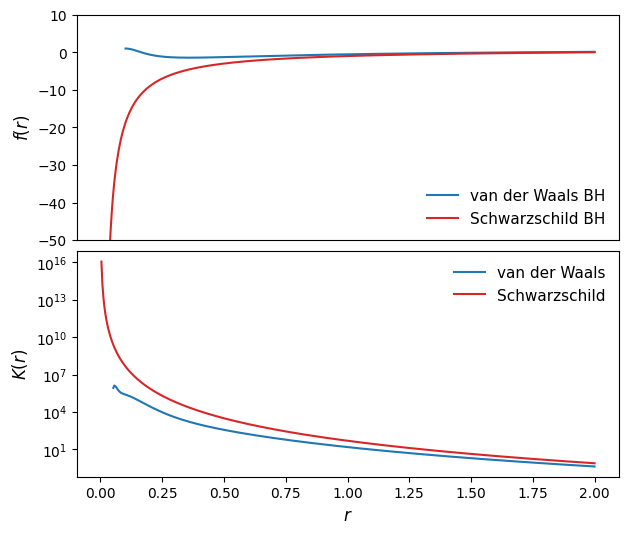}
    \caption{Behaviour of the metric function $f(r)$ and the Kretschmann scalar.}
    \label{fig:2}
\end{figure}

The spatial embedding diagram of this solution, compared with the Schwarzschild case, is illustrated in Fig. \ref{fig:1}. Notably, at $r = r_0$, the geometry becomes locally flat, reflecting the effective decoupling of gravity at short distances.
We further examine the behavior of the metric function \eqref{eq:4.17} and the Kretschmann scalar \eqref{eq:4.12}, as illustrated in Fig. \ref{fig:2}. At large radii, the solution asymptotically reproduces the Schwarzschild geometry, confirming the correct far-field limit. In contrast, as $r\to r_0$, the Schwarzschild solution exhibits divergent curvature invariants, whereas in the present model the Kretschmann scalar remains finite, explicitly demonstrating the resolution of the central singularity and the formation of a regular core. Unlike other regular black hole constructions, such as the Bardeen \cite{Ayon-Beato:2000mjt} or Hayward \cite{Hayward:2005gi} solutions, where singularity resolution relies on the introduction of additional fields, here the regularization arises purely from a geometric mechanism encoded in the scale dependence of the effective gravitational coupling. Consequently, this framework illustrates that the elimination of the black hole singularity can be achieved directly through the underlying geometry, without invoking extra degrees of freedom.

\begin{figure}[!h]
    \includegraphics[width=0.45\textwidth]{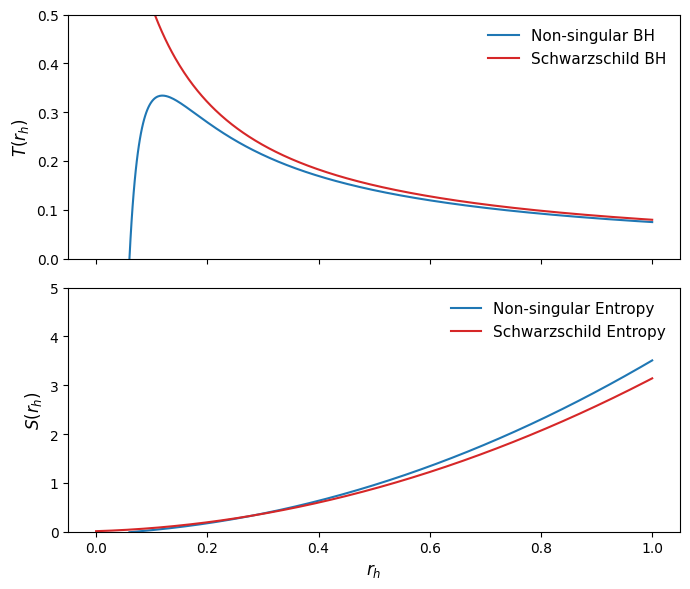}
    \caption{Temperature and entropy as functions of the horizon radius, compared with the Schwarzschild case.}
    \label{fig:3}
\end{figure}

An important consequence of this modification concerns black hole thermodynamics. The presence of the geometric ultraviolet cutoff $r_0$ induces deviations from the Schwarzschild case. Computing the surface gravity $\kappa$, one obtains the modified Hawking temperature
\begin{equation}\label{eq:4.18}
    T = \frac{1}{8\pi}\left(\frac{7}{r_h + r_0} - \frac{5}{r_h - r_0}\right),
\end{equation}
where $r_h$ denotes the horizon radius. The corresponding entropy is given by
\begin{equation}\label{eq:4.19}
\begin{aligned}
    S &=  \frac{\pi \sqrt{r_h^2 - r_0^2}}{3G (r_h - r_0)^2} \left(3r_h^3 + 30r_h^2 r_0 - 229 r_h r_0^2 + 164 r_0^3\right) \\ &\quad +\frac{35\pi r_0^2}{G}\, \ln\left(r_h + \sqrt{r_h^2 - r_0^2}\right).
\end{aligned}
\end{equation}
In the limit $r_0 \to 0$, one recovers the standard Schwarzschild results,
\begin{equation}\label{eq:4.20}
    T_{\mathrm{Sch}} = \frac{1}{4\pi r_h}, 
    \qquad 
    S_{\mathrm{Sch}} = \frac{\pi r_h^2}{G}.
\end{equation}
The evolution of the temperature and entropy is shown in Fig. \ref{fig:3}. A notable deviation from the Schwarzschild spacetime arises in the regime $r_h \sim r_0$. From Eqs. \eqref{eq:4.18} and \eqref{eq:4.19}, the requirement of positive temperature and entropy imposes the condition
\begin{equation}\label{eq:4.21}
    r_h > 6 r_0,
\end{equation}
indicating that quantum-scale black holes can only exist for event horizons satisfying Eq. \eqref{eq:4.21}. Assuming, for instance, that the event horizon is quantized as $r_h=\gamma r_0$, where $\gamma$ is a natural number greater than unity, the temperature and entropy of the regular black hole take the form
\begin{equation}\label{eq:4.22}
    T = \frac{\gamma - 6}{4 \pi r_0 (\gamma^2 - 1)},
\end{equation}

\begin{equation}\label{eq:4.23}
\begin{aligned}
    S &= \frac{\pi r_0^2 \sqrt{\gamma^2 - 1} \left(3 \gamma^3 + 30 \gamma^2 - 229 \gamma + 164 \right)}{3 G (\gamma + 1)(\gamma - 1)^2} \\ &\quad + \frac{35 \pi r_0^2 \ln \left[ r_0 (\gamma + \sqrt{\gamma^2 - 1}) \right]}{G (\gamma + 1)} \, .
\end{aligned}
\end{equation}
Analysis of these expressions shows that both quantities are positive whenever condition \eqref{eq:4.21} is satisfied. For example, setting $\gamma=7$ yields
\begin{equation}\label{eq:4.24}
    T \approx 0.0052, \quad\quad 
    S \approx \frac{\pi r_0^2}{G} \left( 20.03 + 4.375 \ln r_0 \right).
\end{equation}
Consequently, the model predicts that the minimum temperature and entropy of a quantum-scale black hole cannot fall below these values. This feature is naturally connected to quantum mechanics, where a nonzero ground-state energy implies that the thermodynamic quantities of the system, such as temperature and entropy, cannot vanish. In comparison with the Schwarzschild black hole, the regular black hole suggests that at quantum scales the temperature remains bounded, while both temperature and entropy are finite and nonzero.

\section{Conclusions}

A fundamental understanding of nature requires a coherent synthesis of gravitational and quantum principles. In this context, purely phenomenological extensions of general relativity, while often successful at describing observational data, particularly in the dark sector, remain conceptually limited. Such approaches may capture surface-level behavior, but they do not address the underlying structure of spacetime. A deeper theoretical framework is therefore essential.

A promising route toward this synthesis is provided by thermodynamics. The intimate connection between gravity, quantum theory, and thermodynamics has been increasingly evident, particularly since the seminal work of Jacobson, who demonstrated that Einstein’s equations can be derived as an equation of state. Thermodynamics offers a natural language for describing energy, its flow, and its transformations, concepts that lie at the heart of both general relativity and quantum mechanics. Moreover, its well-established foundation within quantum theory makes it an ideal framework for bridging these domains.

Motivated by this perspective, we have revisited the Clausius relation by replacing the ideal gas equation of state with the van der Waals form, thereby incorporating more realistic microscopic interactions. Following Jacobson’s approach, this modification naturally leads to a generalized theory of gravity, in which the gravitational coupling depends on a geometric variable and can evolve according to the properties of the horizon surface. This extension not only defines a new framework for gravity but also establishes a close connection with the holographic principle, quantum mechanics, and black hole thermodynamics. To assess its implications, we have briefly explored both cosmological and black hole solutions.

In the cosmological context, analysis of the modified first Friedmann equation shows that, in the early Universe, these corrections not only resolve the initial singularity but also provide a mechanism for inflation. At late times, the theory reproduces a Friedmann equation analogous to that of loop quantum cosmology, suggesting a consistent description across cosmic epochs.

For static, spherically symmetric solutions, examining the evolution of the effective gravitational coupling allows the construction of a new regular black hole solution. In the asymptotic regime $r\gg1$, it reproduces the Schwarzschild behavior, whereas near $r \sim r_0$, the model avoids singularities. Thermodynamic analysis of the event horizon indicates that quantum-scale black holes can exist provided $r_h >6 r_0$. This condition not only removes the divergent temperature of the Schwarzschild solution near the singularity but also aligns with the nonvanishing ground-state energy, which can be interpreted as the minimal, nonzero contribution to the system’s thermodynamic parameters.

The framework developed in this work provides a geometrically grounded extension of general relativity that incorporates key features expected from quantum gravity. While the present analysis has focused on its foundational aspects, a comprehensive exploration of its consequences remains to be carried out. In particular, further investigations in astrophysical and cosmological contexts, as well as a deeper examination of its relation to quantum theory, are necessary to fully assess its viability. We hope that this work will stimulate further studies in this direction.

\section*{ACKNOWLEDGMENT}
I would like to thank A. H. Fazlollahi for his invaluable support and for nurturing the development of the ideas presented in this work.

\end{document}